\documentclass[aps,prc,onecolumn,showpacs,superscriptaddress]{revtex4-1}
\usepackage{epsfig}
\usepackage{graphicx}
\usepackage{amsmath,amssymb,amsfonts}
\usepackage{array}
\usepackage{url}
\usepackage{hyperref}
\usepackage{multirow}
\usepackage{float}
\usepackage{lineno}
\usepackage{xspace}
\usepackage{float}

\usepackage[usenames,dvipsnames]{color}

\newcommand{\bef}{\begin{figure}}
\newcommand{\eef}{\end{figure}}
\newcommand{\bc}{\begin{center}}
\newcommand{\ec}{\end{center}}

\newcommand{\be}{\begin{equation}}
\newcommand{\ee}{\end{equation}}
\newcommand{\bea}{\begin{eqnarray}}
\newcommand{\eea}{\end{eqnarray}}

\begin{document}

\title{$J/\psi$ production as a function of charged-particle multiplicity with ALICE at the LHC }
\author{Dhananjaya Thakur for the ALICE Collaboration}
\affiliation{Discipline of Physics, School of Basic Sciences, Indian Institute of Technology Indore, Indore- 453552, INDIA}
\email{Dhananjaya.Thakur@cern.ch}
\begin{abstract}
At LHC energies, the charged-particle multiplicity dependence of particle production is a topic of considerable interest in $pp$ collisions. It has been argued that multiple partonic interactions play an important role in particle production mechanisms, not only affecting the soft processes but also the hard processes. Recently, ALICE has measured $J/\psi$ production as a function of charged-particle multiplicity to study the correlation between soft and hard processes. In this contribution, we present the $J/\psi$ production versus multiplicity for $pp$ and $p$\textendash$Pb$ collisions measured by ALICE. We compare the results with different theoretical models.
 \end{abstract}
\date{\today}
\maketitle
\section{Introduction}
\label{intro}
Understanding the mechanism of charmonium production is one of the major challenges in $pp$ collisions.  There are many theoretical  models that try to explain heavy-flavor production in hard scattering processes.  Examples are the Color Singlet Model~\cite{Einhorn:1975ua}, non-relativistic QCD (NRQCD)~\cite{Bodwin:1994jh} and the Color Evaporation Model~\cite{Halzen:1977rs}. The production of charm and anti-charm quark pairs are described by perturbative Quantum Chromodynamics (pQCD) and their binding into  charmonium states by non-perturbative QCD. 
Charmonium suppression is a universally accepted probe for the de-confined medium in heavy-ion collisions. To understand the suppression, it is necessary to understand $J/\psi$ production in $pp$ collisions and also potential cold nuclear matter effects in $p$--$Pb$ collisions. Recently, ALICE has measured $J/\psi$ as a function of charged-particle multiplicity in $J/\psi \rightarrow \mu^{+}  \mu^{-}$ and $J/\psi \rightarrow e^{+}  e^{-}$ at $\sqrt{s}$ = 7 TeV, and observed an increasing trend with respect to charged-particle multiplicity~\cite{Abelev:2012rz}. A similar result has been found for D-meson production. This reveals that the multiple partonic interaction (MPI) which was thought to affect only soft processes can also affect the hard processes and hence $J/\psi$ production.  With the help of new data at $\sqrt{s}$ = 13 TeV, we can measure the trend with multiplicity more precisely than what was previously possible at $\sqrt{s}$ = 7 TeV.  Therefore, to have a clear view of the observed picture, ALICE has extended this analysis to $pp$ collisions at $\sqrt{s}$ = 13 TeV and $p$--$Pb$ collisions at $\sqrt{s_{NN}}$ = 5.02 TeV. We present the results of this analysis and compare them to available theoretical models.
\section{Experimental setup and analysis procedure}
\label{sec:expt}
ALICE is one of the four major experiments at the LHC. Details about ALICE can be found in Ref.~\cite{Aamodt:2008zz}. For the present work, two different spectrometers have been used for  $J/\psi$ reconstruction. The central barrel detector covers the rapidity range $|y|<0.9$ and includes the Inner Tracking System (ITS) and Time Projection Chamber (TPC). It is used for the reconstruction of $J/\psi$ via the di-electron decay channel.~The forward muon spectrometer is used for the reconstruction of $J/\psi$ via the di-muon decay channel in the rapidity range -4.0$<y<$-2.5. Two V0 scintillator arrays used for triggering and are located at -3.7 $< \eta <$ -1.7 and at 2.8 $< \eta <$ 5.1. The V0 detectors are also used as a high-multiplicity trigger.

The charged-particle pseudo-rapidity density ($\textrm{d}N_{\textrm{ch}}/\textrm{d}\eta$) is measured at mid-rapidity~($|\eta| < $1.0)~from the information of track segments (tracklets) in the Silicon Pixel Detector (SPD). Several cuts are applied to determine the accurate position of the z-coordinate of the vertex ($z_{vtx}$). Tracklets are measured within  $|\eta| < $1.0 and  $ |z_{vtx}| < $ 10.0 cm. This account for the  SPD acceptance.  A $z_{\textrm{vtx}}-$dependent correction is applied using data driven method~\cite{javier_thesis}. This also take into account the SPD inefficiency and acceptance.~The correction factor is randomized on an event-by-event basis using a Poisson distribution, for the matching of true charged-particle multiplicity and the tracklet multiplicities.

The self-normalized $J/\psi$ yield in bins of charged-particle multiplicity is calculated as:

\begin{equation}
\begin{aligned}
\label{eq:jpsi_relativeyield}
\frac{\frac{\textrm{d}N_{\textrm{J}/\psi}}{\textrm{d}y}}{\langle \frac{\textrm{d}N_{\textrm{J}/\psi}}{\textrm{d}y}  \rangle} = \frac{N_{\textrm{J}/\psi}^{\textrm{corr},~\textrm{i}}}{N_{\textrm{J}/\psi}^{\textrm{corr},~\textrm{integrated}}} \times \frac{N_{\textrm{MB}}^{\textrm{integrated}}}{N_{\textrm{MB}}^{\textrm{i}}},
\end{aligned}
\end{equation} 
where ($N_{\textrm{J}/\psi}^{\textrm{corr}, \textrm{i}}$, $N_{\textrm{J}/\textrm{$\psi$}}^{\textrm{corr}, \textrm{integrated}}$)  and   ($N_{\textrm{MB}}^{i}$, $N_{\textrm{MB}}^{integrated}$) are the corrected number of $J/\psi$ and number of minimum bias events in $i^{th}$ multiplicity bin and integrated  over all multiplicity bins, respectively.
In $p$--$Pb$ collisions, the mean $p_{\textrm{T}}$ of $J/\psi$, $\langle p_{\textrm{T}}^{J/\psi} \rangle$, is found by fitting the mean $p_{\textrm{T}}$ of unlike-sign pairs of muons as a function of the dimuon invariant mass.

\section{Results}
\label{sec:result}

\begin{figure*} [h]
\begin{center}
\includegraphics[width=14.5pc]{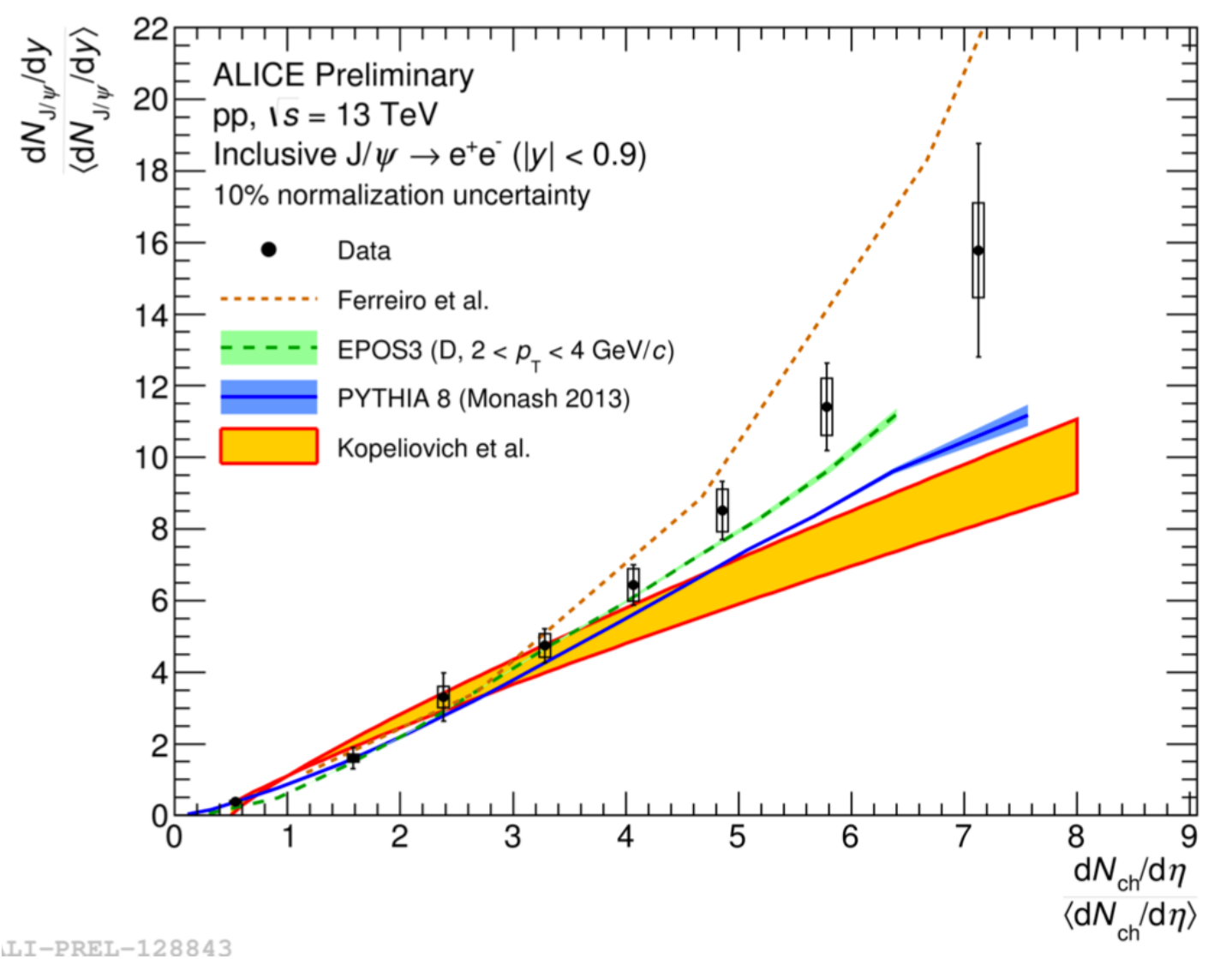}
\includegraphics[width=15.5pc]{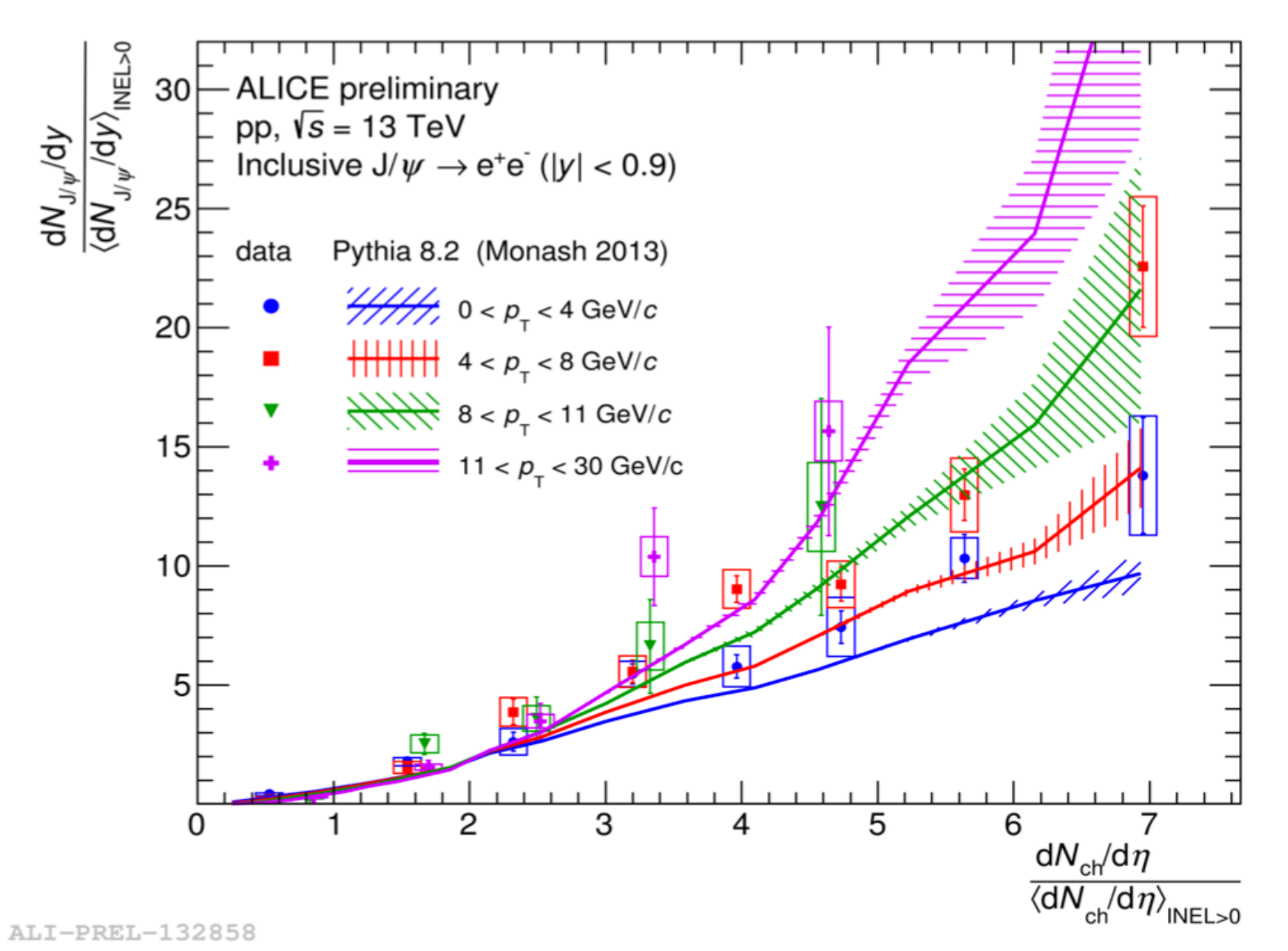}
\caption{Self-Normalized yield of inclusive $J/\psi$ production as a function of multiplicity at mid-rapidity for $pp$ collisions at $\sqrt{s}$ = 13 TeV. The result for integrated transverse momentum compared with predictions from different models (left panel) and result in transverse momentum slices compared with PYTHIA~8 (right panel). }
\label{fig1}
\end{center}
\end{figure*}

Figure \ref{fig1} shows the relative yield of inclusive $J/\psi$ at mid-rapidity as a function of charged-particle multiplicity for integrated $p_{\textrm{T}}$ and in $p_{\textrm{T}}$ slices. It can be see from the figure that a stronger than linear increase of yield is observed as compared to the charged-particle multiplicity. These data are more precise and extend the multiplicity reach with respect to the results at $\sqrt{s}$ = 7 TeV. The result is compared to four models: Ferreiro et al.~\cite{Ferreiro:2012fb}, EPOS3~\cite{Drescher:2000ha}, PYTHIA~8~\cite{Sjostrand:2007gs}~and~Kopeliovich et al.~\cite{Kopeliovich:2013yfa}.

\begin{itemize}
 \item The model of \textbf{Ferreiro et al.}, which is able to well-explain $pp$ result at $\sqrt{s}$ = 7 TeV \cite{Ferreiro:2012fb}, overestimates the $J/\psi$ yield at high multiplicities in $pp$ at $\sqrt{s}$ = 13 TeV. This model assumes that in high energy hadronic collisions all the interacting partons have finite spatial extension and thus collide at finite impact parameter by means of parton-parton collisions. It considers $color~strings$ as the fundamental degrees of freedom. According to this model, $J/\psi$ multiplicity is proportional to number of strings produced ($N_{\textrm{s}}$), whereas charged-particle multiplicity behaves roughly as $\sqrt{N_{\textrm{s}}}$, due to the interaction among the strings.

 \item \textbf{EPOS3} includes MPI and hydrodynamical expansion of the system, describes well the azimuthal correlation of D-meson with charged-particle\cite{ALICE:2016clc}, is also describes the multiplicity dependence of $J/\psi$ production. The good agreement of the EPOS3 model with data shows that the energy density reached in $pp$ collisions at the LHC might be high enough to be described by a hydrodynamical evolution.
 
\item \textbf{PYTHIA~8} has MPI and color reconnection in the final state underestimates the data towards the higher multiplicity bins.

\item The model of \textbf{Kopeliovich et al.} assumes higher Fock states in protons, which contain increased number of gluons. Inelastic collisions of the Fock components lead to high hadron multiplicity and the relative production of $J/\psi$ is enhanced in such gluon-rich collisions.
 
  \end{itemize}

All the models containing MPI, qualitatively reproduce the multiplicity dependence of $J/\psi$ production, which reveals the importance of MPI in $pp$ collisions and in particular for heavy-flavor production. Among all, EPOS is describing the data best. The self normalized $J/\psi$ yield as a function of charged-particle multiplicity~studied~in four $p_{\textrm{T}}$ intervals is shown in the right panel of Fig.\ref{fig1}. The results are compared with PYTHIA8. It reproduces the multiplicity and $p_{\textrm{T}}$ dependence well, further highlighting the importance of MPI. The enhancement is strongest at high $p_{\textrm{T}}$, indicating that the effect of MPI is more important at higher $p_{\textrm{T}}$ in the production of $J/\psi$.

\begin{figure*} [h]
\begin{center}
\includegraphics[width=14.5pc]{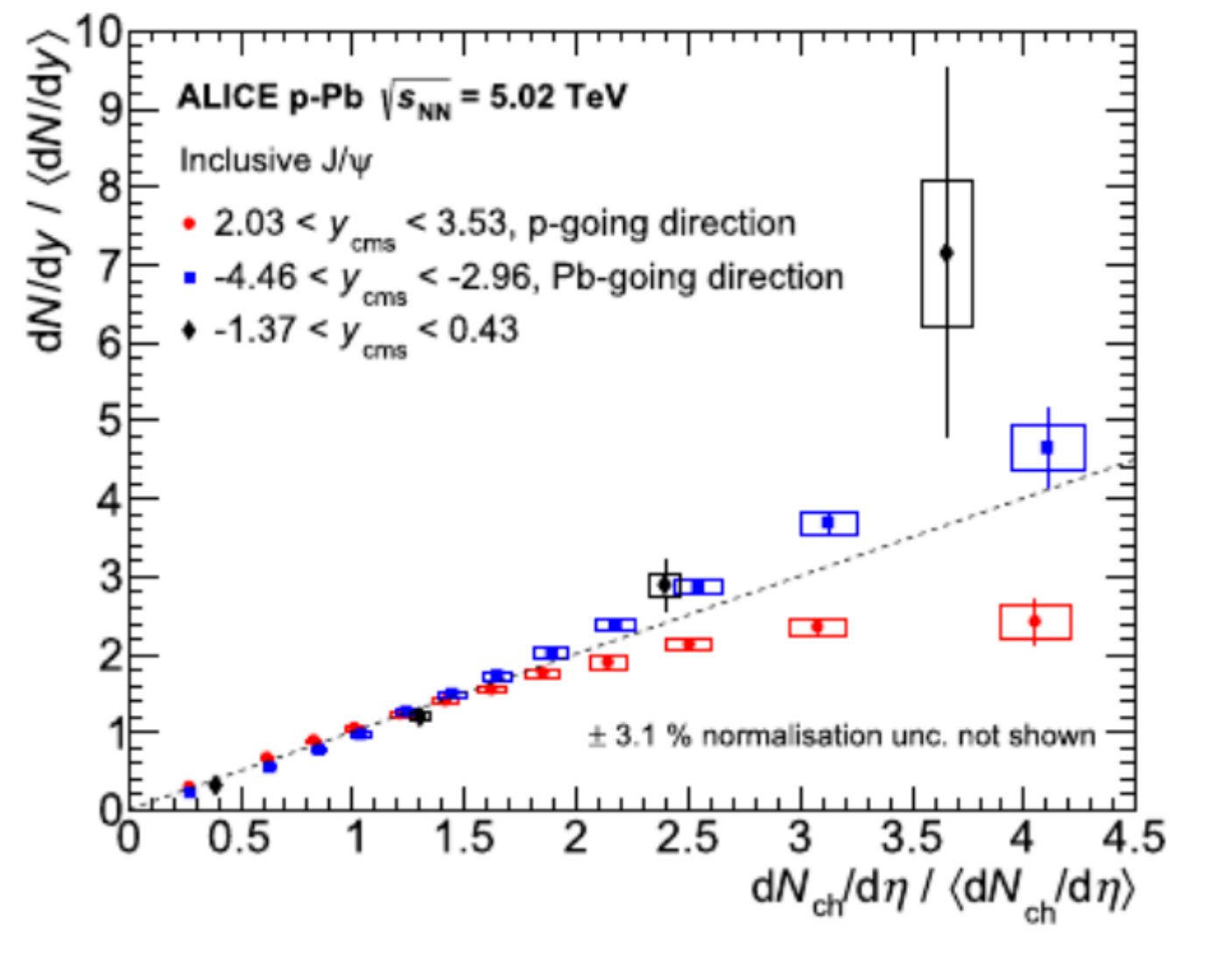}
\includegraphics[width=15pc]{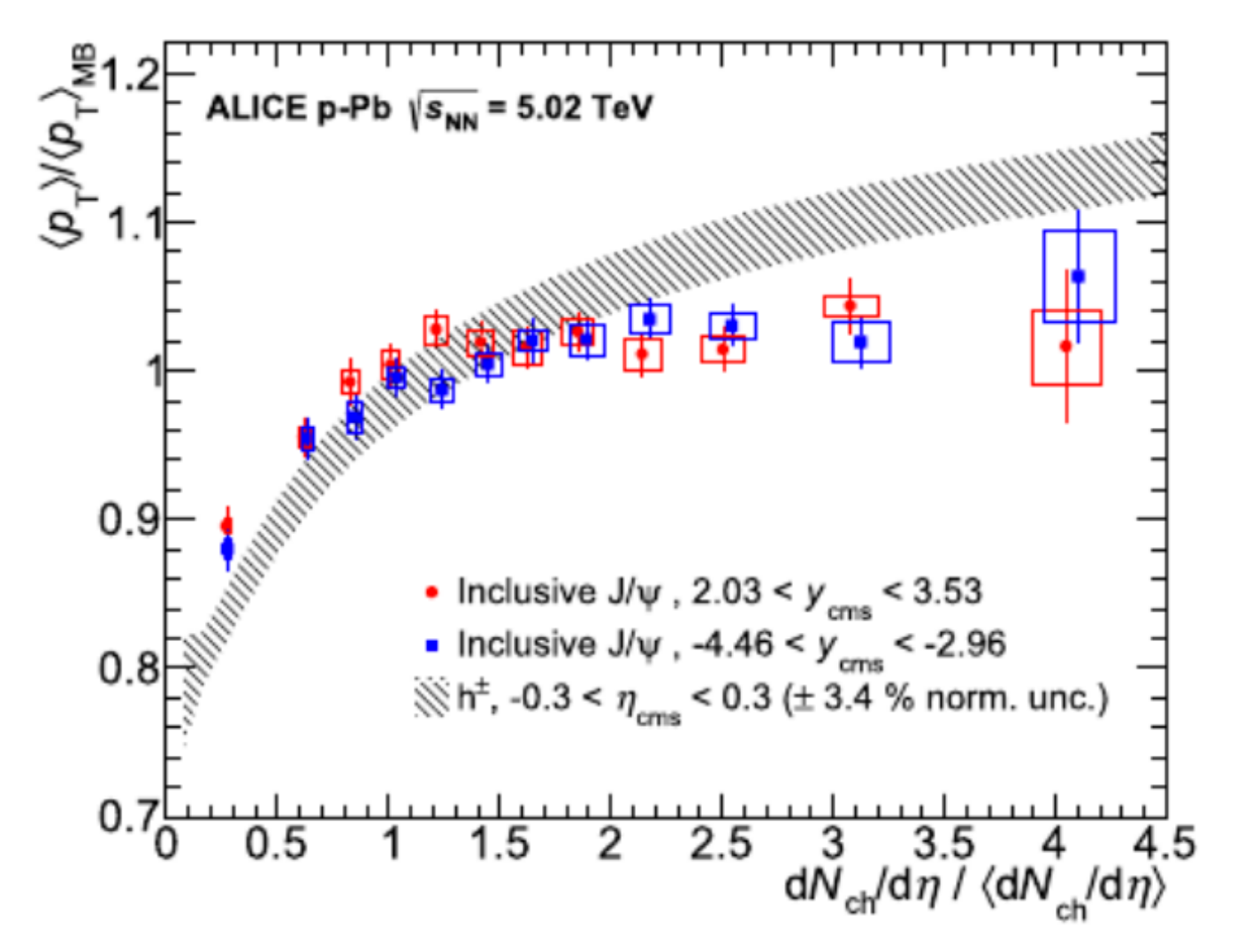}
\caption{Self-normalized yield and mean transverse momentum of $J/\psi$ as a function of self-normalized charged-particle multiplicity for $p$--$Pb$ collisions at $\sqrt{s_{NN}}$ = 5.02 TeV. The result for self-normalized yield of inclusive $J/\psi$ production at forward ($2.03 < y_{\textrm{cms}} < 3.53$), backward ($-4.46 < y_{\textrm{cms}} < -2.96$) and mid-rapidity ($-1.37< y_{\textrm{\textrm{cms}}} < 0.43$) are shown in left pannel. The result for self-normalized mean transverse momentum of $J/\psi$ at forward ($2.03 < y_{\textrm{cms}} < 3.53$) and backward rapidity ($-4.46 < y_{\textrm{cms}} < -2.96$) are shown in right panel.}  
\label{fig2}
\end{center}
\end{figure*}

The multiplicity dependence of~$J/\psi$ production as a function of charged-particle multiplicity has also been studied in $p$--$Pb$ collisions at           
 $\sqrt{s_{NN}}$ = 5.02 TeV \cite{Adamova:2017uhu}. ALICE has performed the study in different rapidity ranges by inverting the directions of the lead and proton beams. The measurement has been performed in three rapidity regions, forward ($2.03 < y_{\textrm{cms}} < 3.53$), backward ($-4.46 < y_{\textrm{cms}} < -2.96$) and mid-rapidity ($-1.37< y_{\textrm{cms}}<0.43$), as shown in the left panel of Fig.\ref{fig2}. The mid-rapidity and backward-rapidity data show a linear increase of self-normalized $J/\psi$ yield with multiplicity. At forward rapidity a saturation towards higher multiplicity is observed. In this kinematic region, the~proton~probes the small Bjorken-x region of the Pb-nucleus, where cold nuclear effect, gluon shadowing and saturation effects are expected.

ALICE has also measured $\langle p_{T}^{J/\psi} \rangle$~as a function of multiplicity at forward and backward rapidities, which is shown in the right panel of the Fig.\ref{fig2}. It can be seen from the figure that both rapidity regions show a similar trend. The $\langle p_{T}\rangle$~is the same within uncertainty. The $\langle p_{T} \rangle$ of charged hadrons is represented by a dashed band. The~$\langle p_{T}^{J/\psi} \rangle$~shows a similar trend as that observed in $Pb$--$Pb$ collisions for charged particles~\cite{Abelev:2013bla}, possibly hinting at collective effects in $p$--$Pb$ collisions. 

\section{Summary} 
\label{sec:sum}   
In this contribution, the multiplicity dependence of $J/\psi$ production has been presented for $pp$ collisions at $\sqrt{s}$ = 13 TeV and $p$--$Pb$ collisions at $\sqrt{s_{NN}}$ = 5.02 TeV.  Also, $\langle p_{T}^{J/\psi} \rangle$ as function of multiplicity for $p-Pb$ collisions at $\sqrt{s_{NN}}$ = 5.02 TeV has been presented at forward and backward rapidities. Preliminary results of $J/\psi$ production as a function of charged-particle multiplicity for $J/\psi \rightarrow e^{+}  e^{-}$ are presented. 
Similar work at forward rapidity looking at the decay channel $J/\psi \rightarrow \mu^{+}  \mu^{-}$ is ongoing. The $J/\psi$ yield versus charged-multiplicity study for $p$--$Pb$ results will help~to~understand differences in the production mechanisms between $pp$ and $p$--$Pb$ collisions.

{}

\end{document}